\documentclass[useAMS]{mn2e}
\usepackage{graphicx}
\usepackage{array}

\begin{document}

\title[Unusual pulsation spectrum of HS\,0507+0434B]{The unusual pulsation 
spectrum of the cool ZZ Ceti star HS\,0507+0434B}
\author[G. Handler, E. Romero-Colmenero \& M. H. Montgomery]
       {G. Handler$^1$, E. Romero-Colmenero$^1$, M. H. Montgomery$^2$ 
\and \\
$^1$ South African Astronomical Observatory, P.O. Box 9, Observatory 7935,
South Africa\\
$^2$ Institute for Astronomy, Madingley Road, Cambridge, CB3 0HA, England,
United Kingdom}
      
\date{Accepted 2002 April 29}

\maketitle

\begin{abstract}

We present the analysis of one week of single-site high-speed CCD
photometric observations of the cool ZZ Ceti star HS\,0507+0434B. Ten
independent frequencies are detected in the star's light variations: one
singlet and three nearly-equally spaced triplets. We argue that these
triplets are due to rotationally split modes of spherical degree $\ell=1$.
This is the first detection of consistent multiplet structure in the
amplitude spectrum of a cool ZZ Ceti star and it allows us to determine
the star's rotation period: 1.70$\pm$0.11\,d.

We report exactly equal {\it frequency}, not period, spacings between the
detected mode groups. In addition, certain pairs of modes from the four
principal groups have frequency ratios which are very close to 3:4 or 4:5;
while these ratios are nearly exact (within one part in $10^{4}$), they
still lie outside the computed error bars. We speculate that these
relationships between different frequencies could be caused by resonances.
One of the three triplets may not be constant in amplitude and/or
frequency.

We compare our frequency solution for the combination frequencies (of
which we detected 38) to Wu's (1998, 2001) model thereof. We obtain
consistent results when trying to infer the star's convective thermal time
and the inclination angle of its rotational axis. Theoretical
combination-frequency amplitude spectra also resemble those of the
observations well, and direct theoretical predictions of the observed
second-order light-curve distortions were also reasonably successful
assuming the three triplets are due to $\ell=1$ modes. Attempts to
reproduce the observed combination frequencies adopting all possible
$\ell=2$ identifications for the triplets did not provide similarly 
consistent results, supporting their identification with $\ell=1$.

\end{abstract}


{\section{Introduction}}

Asteroseismology is the only observational method which permits the
exploration of the deep interior structure of stars. Just as we know the
Earth's interior from the analysis of earthquakes, the oscillations of
pulsating stars are used to probe stellar interiors. Pulsating white dwarf
stars are well-suited for asteroseismology, as they often show a large
number of excited normal modes, each of them carrying a different part of
the information. Asteroseismology has therefore been quite successful for
some of these objects. The interior structures of the DO star PG 1159-035
(Winget et al. 1991) and the DB white dwarf GD 358 (Winget et al. 1994)
were constrained with unprecedented accuracy.

However, this task proved to be more difficult for the coolest known white
dwarf pulsators of spectral type DA, the ZZ Ceti or DAV stars (see
Kleinman 1999 for a review). They can be separated into two groups, hot
and cool ZZ Ceti stars, which show different pulsation properties. The hot
ZZ Ceti stars have only a few excited modes of short period (2 -- 5 min)
and low amplitude ($<$ 0.05 mag), and can, in principle, be understood as
a group (Clemens 1994). The cool ZZ Ceti stars, on the other hand, have
higher amplitudes ($>$ 0.1 mag) and longer periods ($\approx$ 10 min) than
their hotter counterparts and their pulsation spectra are generally
variable in time. This, together with their few independent pulsation
modes, left them as a mystery for a long time.

The first cool ZZ Ceti star which can be regarded as reasonably well
understood is G29-38. Kleinman et al. (1998) finally managed to decipher
its pulsational mode structure after 24 different monthly observing runs,
including two Whole Earth Telescope (WET, Nather et al. 1990) campaigns!
Consequently, it seems that large amounts of data are necessary to
understand the pulsations of such stars and, therefore, few cool ZZ Ceti
stars are well observed (given their complexity) in terms of time-series
photometry.

In the case of HS\,0507+0434B, the secondary component in a double
degenerate DA binary system (Jordan et al. 1998), only a three-hour
discovery light curve demonstrating the presence of multi-periodic
pulsations was available. The analysis of these data showed clear signs of
the pulsations of a cool ZZ Ceti star: a few pulsation modes with a number
of combination frequencies. However, as HS\,0507+0434B is a member of a
wide binary system where both components can be assumed to have evolved as
single stars, further constraints aiding its modelling can be utilized
(see Jordan et al. 1998), which makes HS\,0507+0434B of increased
interest. Consequently, we decided to study the star in more detail to
explore its asteroseismological potential.

\vspace{4mm}

\section{Observations and reductions}

HS\,0507+0434B was observed with the University of Cape Town CCD camera
(O'Donoghue 1995) on the 0.75-m telescope at the Sutherland station of the
South African Astronomical Observatory for seven consecutive nights in
January 2000. The instrument was operated in frame-transfer mode, which
allows continuous integrations on the field of choice; integration times
of 10 seconds were used. HS\,0507+0434B ($V = 15.36$) and three comparison
stars (including HS\,0507+0434A, a hotter DA white dwarf) of similar
brightness were observed. No filter was used. In every clear night, sky
flat fields were taken during twilight. The journal of our observations is
presented in Table\,1.

\begin{table}
\caption[]{Time series photometry of HS\,0507+0434B}
\begin{center}
\begin{tabular}{cccc}
\hline
Run No. & Date & Start & Run length \\
 & (UT) & (UT) & (h) \\
\hline
1 & 4 Jan 2000 & 22:20:02 & 2.61 \\
2 & 5 Jan 2000 & 20:00:37 & 4.65 \\
3 & 6 Jan 2000 & 20:54:23 & 3.60 \\
4 & 7 Jan 2000 & 19:12:53 & 4.52 \\
5 & 8 Jan 2000 & 19:13:16 & 5.31 \\
6 & 9 Jan 2000 & 18:52:58 & 5.32 \\
7 & 10 Jan 2000 & 18:39:15 & 6.04 \\
\hline
Total & & & 32.05 \\
\hline
\end{tabular}
\end{center}
\end{table}

Data reduction was started with correction for bias and flat field. After
examining the nightly mean flat field images for possible variations (and
not finding any), we formed a mean flat field for the whole run by which
each science frame was divided. The stellar magnitudes were determined
through PSF-fitting with a modified version (O'Donoghue, private
communication) of DoPhot (Schechter, Mateo \& Saha 1993). No evidence for
variability of any of the comparison stars was found.

We then corrected for differential colour extinction ($k_C\approx 0.04$)
indicating that the target is bluer than the ensemble of comparison stars.
A correlation of the differential magnitudes of HS\,0507+0434B with seeing
was not found.

Finally, some low-frequency trends in the data were removed by means of
low-order polynomials, and the times of measurement were transformed to a
homogeneous time base. Terrestrial Time (TT) served as our reference for
measurements on the Earth's surface and a barycentric correction was
calculated. As this barycentric correction varied by about $-$1\,s
throughout a typical run, we applied it point by point. Thus our final
time base is Barycentric Julian Ephemeris Date (BJED). The final time
series was subjected to frequency analysis. Fig.\,1 shows an example
light curve.

\begin{figure}
\includegraphics[width=99mm,viewport=-50 -30 265 325]{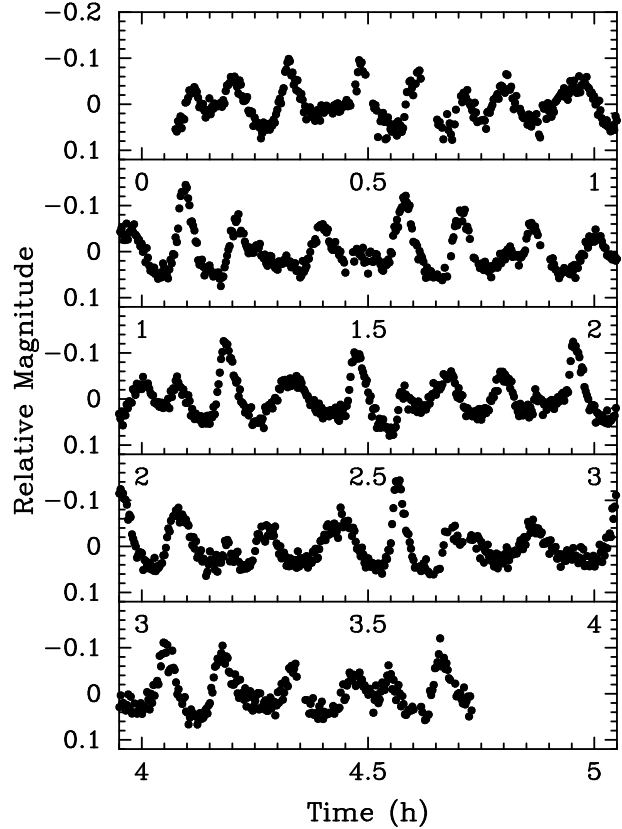}
\caption[]{The reduced light curve of HS\,0507+0434B from January 5, 2000.
Note the multi-periodic variability.}
\end{figure}

\section{Frequency analysis}

Our frequency analysis was performed with the program {\tt Period98}
(Sperl 1998). This package applies single-frequency Fourier analysis and
simultaneous multi-frequency sine-wave fitting. {\tt Period98} can also be
used to calculate optimal fits for multi-periodic signals including
harmonics and combination frequencies at values fixed relative to the
``parent'' modes. Such fixed-frequency solutions increase the stability of
the results due to the decrease of the number of free parameters in the
fit and by its smaller vulnerability to aliasing and noise problems.

\begin{figure}
\includegraphics[width=99mm,viewport=10 00 315 605]{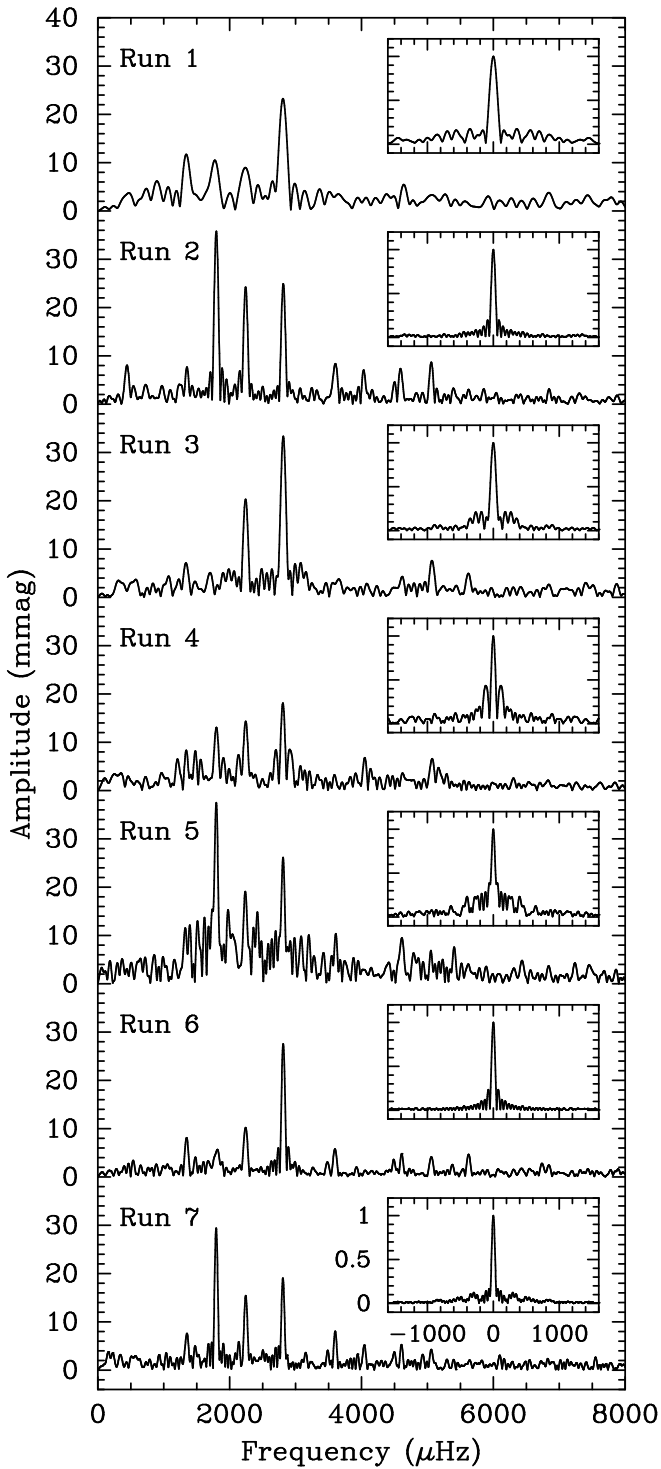}
\caption[]{Nightly amplitude spectra of HS\,0507+0434B; the corresponding
spectral window patterns are shown as insets. There seem to be three
dominant modes (``regions of power'') which are variable in amplitude. The
presence of combination frequencies can also be suspected.}
\end{figure}

We calculated nightly amplitude spectra of our data; they are shown in
Fig.\,2. The amplitude spectra of the individual nights are different, but
the dominant peaks seem to occur at the same frequencies. Their amplitudes
vary however, and there is no obvious correspondence between the
variations of the different peaks. For instance, the peak near 2800
$\mu$Hz appears to vary less in amplitude than the one near 1800 $\mu$Hz.

As the next step, we combined all the different runs and calculated
amplitude spectra for the resulting data set. The full amplitude spectrum
is displayed in Fig.\,3, which contains a number of dominant features and
combination frequencies. Some of the structures are more complicated than
a single frequency convolved with the spectral window. We will now
concentrate on the three strongest features.

\begin{figure*}
\includegraphics[width=99mm,viewport=180 00 665 609]{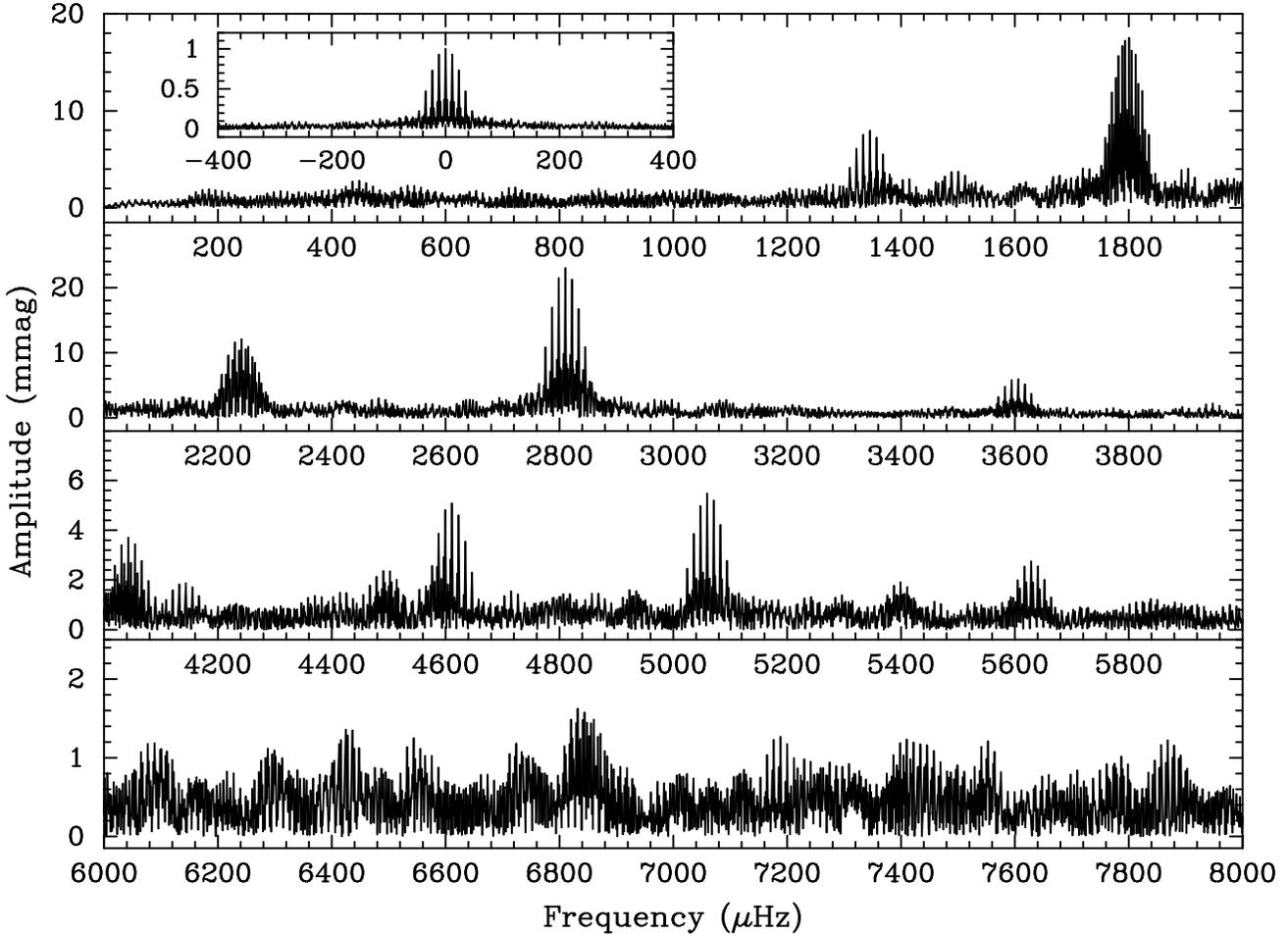}
\caption[]{Amplitude spectrum of our combined HS\,0507+0434B data. The
inset is the spectral window of the data. As our data were taken from a
single site only, aliasing is present. Note the different ordinate scale 
of each panel.}
\end{figure*}

In Fig.\,4, we show prewhitening within these features by simultaneous
optimization of the frequencies, amplitudes and phases of all the detected
signals. We consecutively removed one, two and three signals from each
structure. We adopted the location of the highest peak in each panel as
the next frequency to be included in the following fit; we verified that
this was indeed the best choice by evaluating the residuals. In the lowest
panel of Fig.\,4, schematic amplitude spectra composed of the detected
signals are shown. All three structures are equally spaced triplets with
the central component lowest in amplitude. The frequency spacings within
these triplets are very similar.

\begin{figure}
\includegraphics[width=99mm,viewport=10 00 390 558]{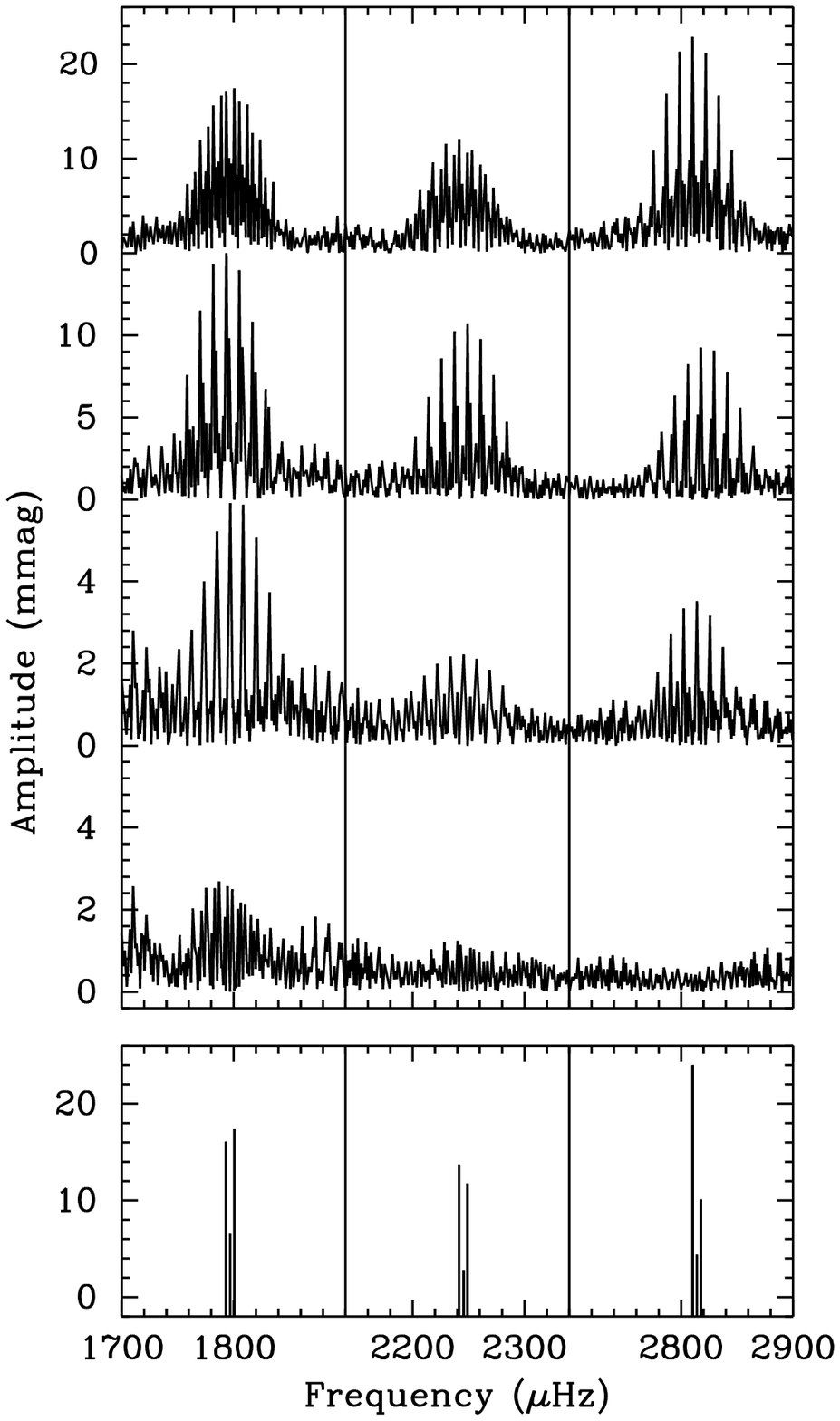}
\caption[]{Amplitude spectra of our HS\,0507+0434B data in the vicinity of
the dominant features. Upper panel: original data, second panel from
top: amplitude spectrum after prewhitening of one frequency in each
section, third panel: after prewhitening of two frequencies each, fourth
panel: three frequencies each prewhitened. The lowest panel shows the
detected frequencies with the corresponding amplitudes schematically. Note
the change of the ordinate scale between the different panels.}
\end{figure}

The residual amplitude spectrum after prewhitening the data with the three
triplets is displayed in the upper panel of Fig.\,5. Several of the
highest peaks were recognized as combinations $m\times f_i \pm n\times
f_j$ of these nine frequencies. Using {\tt Period98} we verified that this
was indeed the case. Consequently, we fixed those frequencies to their
exact value predicted from the parent frequencies. We also suspected
that more combination frequencies were present in the residuals. We
searched for them using the program {\tt lincom} (Kleinman 1995), which
finds all possible combinations automatically.

We detected many more combination frequencies clearly exceeding the noise
level in the amplitude spectrum. For some of these more than one
identification is possible. In such cases, e.g. frequency sums or
differences of different multiplets which are ``degenerate'', we adopted
the most plausible one. This would be the one matching best in frequency, 
the one yielding the lowest residual between light curve and fit or the
one which would be generated by the highest-amplitude parent modes. Thus
we caution that some matches must be regarded as formal, especially where
higher-order combinations are concerned; some of these can have many
different potential combinations of parent modes.

After identification and prewhitening of all detectable combination
frequencies of the components of the three triplets, we calculated the
residual amplitude spectrum (second panel of Fig.\,5). One dominant peak
and a few other signals are visible. Including this highest peak and its
combinations with previously detected signals into the multi-frequency
solution and prewhitening it from the data results in the new residual
amplitude spectrum shown in the third panel of Fig.\,5. Some peaks still
exceed the noise level in this amplitude spectrum considerably; thus their
frequencies and amplitudes were determined as well. The final residual
amplitude spectrum can be found in the lowest panel of Fig.\,5. 

The multi-frequency solution we derived is summarized in Table 2 together
with 1-$\sigma$ error estimates calculated with the formulae by Montgomery
\& O'Donoghue (1999). The detected signals are ordered in the following
way: corresponding to increasing frequency, the independent modes are
listed first, then two-mode frequency sums followed by two-mode frequency
differences, three-mode frequency combinations, then fourth-order
combinations and finally further signals, some with suggested
identifications.

\begin{figure}
\includegraphics[width=99mm,viewport=00 00 503 512]{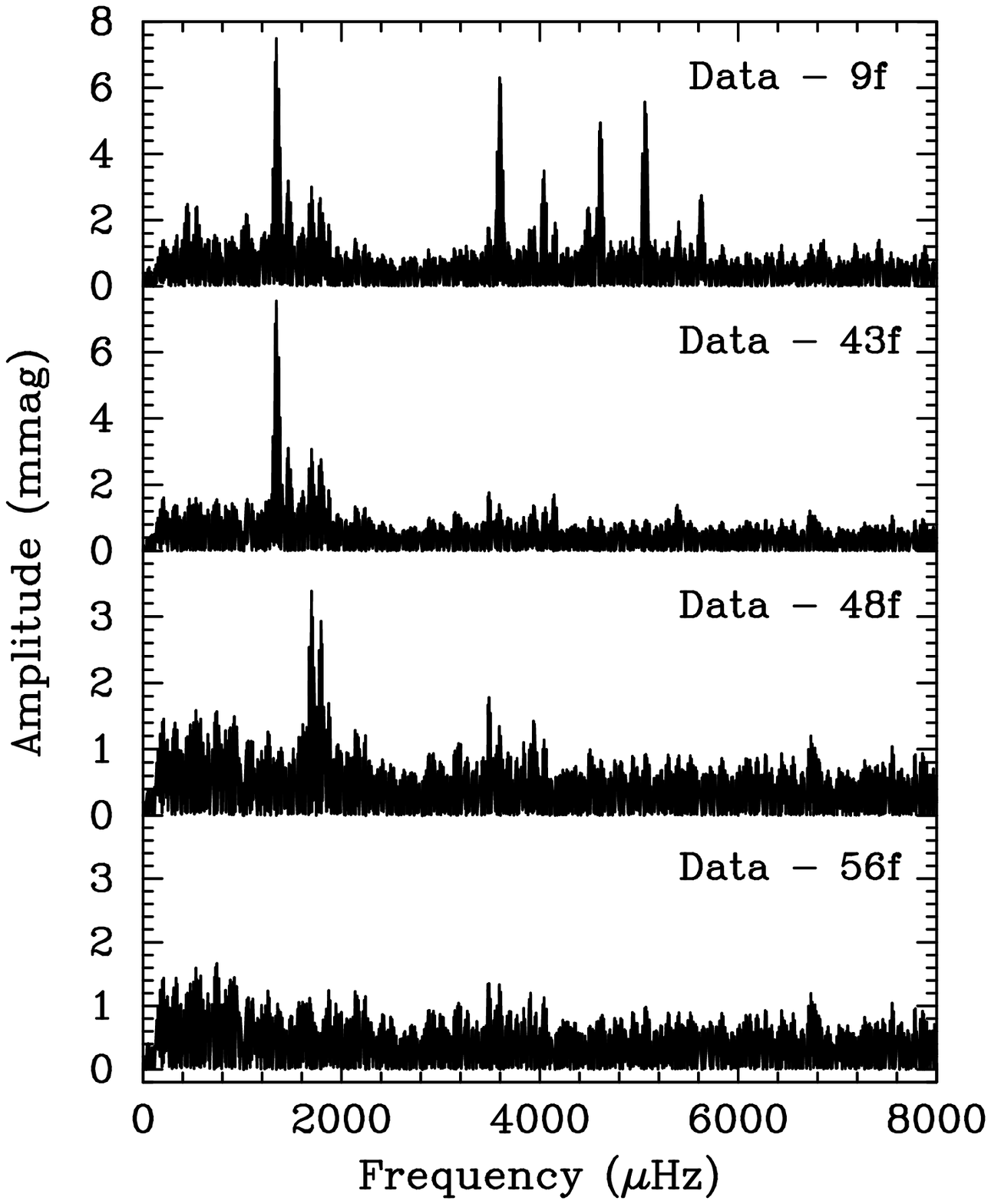}
\caption[]{Residual amplitude spectra of HS\,0507+0434B with consecutive
prewhitening of detected signals. Uppermost panel: residuals after
prewhitening of the three triplets decomposed in Fig.\,4. Second panel:
residuals after prewhitening these triplets as well as their frequency
combinations. Third panel (ordinate scale magnified): residuals after
subtracting one more signal plus its combinations. Lowest panel: residuals
after including seven more (partly suspected) periodicities.}
\end{figure}

\begin{table}
\caption[]{Our multi-frequency solution for HS\,0507+0434B. Individual
error estimates are quoted for the independent frequencies and periods;
the amplitude uncertainty is $\pm 0.2$ mmag. Frequencies are given in
$\mu$Hz, amplitudes in mmag and periods in seconds. Combinations
marked with asterisks were detected outright.}
\begin{center}
\begin{tabular}{cccc}
%\hline
\hline
\multicolumn{4}{c}{{\bf Independent signals}} \\
%\hline
ID & Frequency & Amplitude & Period \\
\hline
$f_1$ & 1345.22 $\pm$ 0.03 & 7.6 & 743.373 $\pm$ 0.017\\
$f_2$ & 1793.31 $\pm$ 0.02 & 16.0 & 557.628 $\pm$ 0.005\\
$f_3$ & 1796.81 $\pm$ 0.03 & 7.2 & 556.542 $\pm$ 0.010\\
$f_4$ & 1800.70 $\pm$ 0.01 & 16.6 & 555.341 $\pm$ 0.004\\
$f_5$ & 2241.47 $\pm$ 0.02 & 13.9 & 446.136 $\pm$ 0.003\\
$f_6$ & 2245.63 $\pm$ 0.08 & 2.8 & 445.309 $\pm$ 0.017\\
$f_7$ & 2249.01 $\pm$ 0.02 & 11.9 & 444.641 $\pm$ 0.004\\
$f_8$ & 2810.37 $\pm$ 0.01 & 24.0 & 355.826 $\pm$ 0.001\\
$f_9$ & 2814.00 $\pm$ 0.05 & 4.3 & 355.366 $\pm$ 0.007\\
$f_{10}$ & 2817.89 $\pm$ 0.02 & 10.2 & 354.875 $\pm$ 0.003\\
\hline
\multicolumn{4}{c}{{\bf Combination frequencies}} \\
%\hline
ID & Combination & Frequency & Amplitude \\
\hline
$f_{11}$ & $f_1+f_4$ & 3145.92 & 0.9 \\
$f_{12}^{\ast}$ & $f_2+f_4$ & 3594.01 & 6.4 \\
$f_{13}$ & $f_3+f_4$ & 3597.51 & 1.1 \\
$f_{14}$ & 2$f_4$ & 3601.40 & 1.6 \\
$f_{15}$ & $f_2+f_5$ & 4034.78 & 1.1 \\
$f_{16}$ & $f_3+f_5$ & 4038.28 & 1.7 \\
$f_{17}^{\ast}$ & $f_4+f_5$ & 4042.17 & 3.3 \\
$f_{18}$ & $f_1+f_8$ & 4155.59 & 1.6 \\
$f_{19}^{\ast}$ & $f_5+f_7$ & 4490.48 & 2.5 \\
$f_{20}$ & $f_2+f_8$ & 4603.68 & 2.6 \\
$f_{21}^{\ast}$ & $f_4+f_8$ & 4611.07 & 5.2 \\
$f_{22}^{\ast}$ & $f_5+f_8$ & 5051.84 & 2.3 \\
$f_{23}^{\ast}$ & $f_7+f_8$ & 5059.37 & 5.9 \\
$f_{24}$ & 2$f_8$ & 5620.73 & 1.8 \\
$f_{25}^{\ast}$ & $f_8+f_{10}$ & 5628.26 & 2.9 \\
$f_{26}$ & 2$f_{10}$ & 5635.79 & 1.0 \\  
$f_{27}$ & $f_5-f_4$ & 440.77 & 2.2 \\
$f_{28}^{\ast}$ & $f_5-f_2$ & 448.16 & 2.4 \\
$f_{29}$ & $f_8-f_5$ & 568.90 & 1.7 \\
$f_{30}$ & $f_8-f_4$ & 1009.67 & 1.3 \\
$f_{31}$ & $f_8-f_2$ & 1017.05 & 2.2 \\
$f_{32}^{\ast}$ & $f_8-f_1$ & 1465.14 & 3.2 \\
$f_{33}$ & $f_3+f_5-f_8$ & 1227.91 & 1.4 \\
$f_{34}$ & $f_7+f_8-f_4$ & 3258.67 & 1.4 \\
$f_{35}$ & $f_1+f_4+f_5$ & 5387.40 & 1.5 \\
$f_{36}$ & $f_2+2f_4$ & 5394.71 & 1.2 \\
$f_{37}$ & 3$f_4$ & 5402.10 & 1.0 \\
$f_{38}$ & $f_3+f_4+f_5$ & 5838.98 & 1.2 \\
$f_{39}$ & $f_3+f_5+f_7$ & 6287.28 & 0.9 \\
$f_{40}$ & $f_8+2f_4$ & 6411.76 & 0.9 \\
$f_{41}$ & $f_4+f_5+f_8$ & 6852.54 & 1.4 \\
$f_{42}$ & $f_3+f_7+f_8$ & 6856.18 & 0.8 \\
$f_{43}$ & $f_4+f_7+f_8$ & 6860.07 & 1.4 \\
$f_{44}$ & $2f_7+f_8$ & 7308.38 & 1.1 \\
$f_{45}$ & $f_4+2f_8$ & 7421.43 & 1.2 \\
$f_{46}$ & $f_5+2f_{10}$ & 7877.26 & 1.0 \\
$f_{47}$ & $2f_2+2f_{10}$ & 6404.52 & 1.0 \\
$f_{48}$ & 3$f_2+f_3$ & 7176.74 & 1.3 \\
\hline
\multicolumn{4}{c}{{\bf Further signals}} \\
%\hline
ID & Frequency & Amplitude & Period/Comb. \\
\hline
$f_{49}$ & 1698.59 $\pm$ 0.07 & 3.6 & 588.73 $\pm$ 0.023\\
$f_{50}$ & 1712.80 $\pm$ 0.16 & 1.5 & 583.84 $\pm$ 0.053\\
$f_{51}$ & 3491.89 & 1.6 & $f_{49}+f_2$ \\
$f_{52}$ & 3495.39 & 1.0 & $f_{49}+f_3$ \\
$f_{53}$ & 3947.59 & 1.4 & $f_{49}+f_7$ \\
$f_{54}$ & 4516.48 & 1.0 & $f_{49}+f_{10}$ \\
$f_{55}$ & 1786.91 $\pm$ 0.09 & 2.5 & 559.62 $\pm$ 0.029\\
$f_{56}$ & 1794.60 $\pm$ 0.08 & 2.9 & 557.23 $\pm$ 0.025\\
%\hline
\hline
\end{tabular}
\end{center}
\end{table}

As mentioned before, the frequency spacing of the three triplets among the
ten ``fundamental'' frequencies is very similar. Consequently, we examined
whether this spacing can be assumed to be exactly the same by
calculating a forced fit with {\tt Period98}, but residual amplitude
spectra then showed conspicuous mounds of power in the frequency domains
of the triplets. This suggests that either the spacing between their
components is not exactly the same, or that (some of) the triplet
components are not constant in amplitude and/or frequency over the whole
data set, or that the light variations require a more complicated
description than the frequency triplets provide.

Our results depend strongly on the correctness of the determination of the
independent frequencies. To check their reliability, we repeated the
frequency analysis, removing single nights from the total data set, thus
restricting ourselves to six nights of data only. We independently
recovered the correct frequencies in all seven possible subsets. Although
light curves of pulsating white dwarf stars are in most cases only
decipherable with observations from many sites, as e.g. used by the WET,
in this special case we think we were fortunate enough to do so with
single-site data only.

\section{Discussion and interpretation}

The discussion of the pulsation spectrum of HS\,0507+0434B is separated
into several sections. Section 4.1 deals with the interpretation of the
ten ``fundamental frequencies'', in Section 4.2 we attempt an
interpretation of the ``further'' signals (cf. Table 2), and in Section
4.3 we will examine the combination frequencies.

\subsection{The ten ``fundamental frequencies''}

\subsubsection{Mode identification}

The pulsation spectrum of HS\,0507+0434B is reminiscent of a cool
ZZ Ceti star: there are a few independent frequencies and many
combinations of these. A very important finding based on the ten
``fundamental'' frequencies is that three equally spaced triplets are
among them. Moreover, all three triplets have the same basic structure in
amplitude: the central component is the weakest.

We suggest that these three independent frequency triplets are caused by
non-radial gravity (g) mode pulsations: the appearance and complexity of
the pulsation spectrum leaves pulsation as the only explanation for the
light variations. Since the fundamental radial mode of a ZZ Ceti star
would have a period of only a few seconds, but the observed time scales of
the pulsations are much longer, they must be due to g-modes, which are
normal-mode pulsations.

We now attempt to constrain the spherical degree $\ell$ of the pulsations.
Modes of $\ell > 2$ are not expected to be detectable due to geometrical
cancellation of the variations over the visible stellar disk (Dziembowski
1977). We are therefore left with the possibilities of modes of $\ell = 1$
only, $\ell = 2$ only, or a mixture of these. We restrict the following
discussion to the three nearly equally spaced triplets, as we have no
means to constrain the $\ell$ value of $f_1$ at this point.

For the three triplets, an explanation in terms of $\ell = 2$ modes only
is unlikely, as one would need to invoke some mode selection mechanism, a
specific inclination of the stellar pulsation axis to the line of sight,
or chance, to produce the more or less consistent triplet structure
observed.

A more natural explanation of the three triplets is that they are all due
to $\ell = 1$ modes. The frequency splitting within the triplets is very
similar, and no special effects need to be assumed to produce such
structures with $\ell = 1$ modes.

The presence of a mixture of $\ell = 1$ and $\ell = 2$ cannot conclusively
be ruled out, although it would be contrived. The different amount of
rotational splitting (with a ratio of about 0.6, see Winget et al. 1991)
within mode groups of $\ell = 1$ and $\ell = 2$ would be immediately
obvious, unless single modes intruding into multiplets of the other $\ell$
would had just the right frequency to fit into the pattern. We consider
this unlikely. In any case, if a mixture of both spherical degrees was
observed, it is more likely that most of the detected frequencies are due
to $\ell = 1$ modes because of the arguments given before.

Consequently, we believe that the most likely mode identification for the
observed triplets in the amplitude spectra of HS\,0507+0434B is that they
are all caused by rotationally split modes of spherical degree $\ell = 1$.
From experience with other ZZ Ceti stars we must however be somewhat
cautious with arguments based on this identification.

\subsubsection{Pulsation periods}

Kleinman (1999) summarized the preferred pulsation periods of ZZ Ceti
stars, with particular focus on the cooler variables. He found concise
groupings of modes from 350 to 650\,s, with a typical spacing of 50\,s.
Our three presumed $\ell = 1$ triplets fit this pattern nicely (they are
close to the ``preferred'' periods of 359, 454 and 546 seconds), which
supports our mode identification further. Conversely, if the pulsation
modes of HS\,0507+0434B were not predominantly $\ell = 1$, this would
represent a challenge for Kleinman's (1999) identifications (if
HS\,0507+0434B is a normal ZZ Ceti star), which we however believe to be
on solid grounds.

Using the periods listed by Kleinman (1999) as a template, we note that
there are gaps within the frequencies of the triplets of HS\,0507+0434B
where additional $\ell = 1$ modes would be expected, suggesting that some
possible radial overtones are not excited to detectable amplitudes. In
addition, the 743-s singlet falls into a period region where no clear
groupings of modes are observed in the ensemble of cool ZZ Ceti stars
investigated by Kleinman (1999).

The gaps in the suspected $\ell=1$ mode pattern of HS\,0507+0434B,
combined with effects of mode trapping in ZZ Ceti star pulsation models
(Brassard et al. 1992b), also preclude an asteroseismological mass
estimate based on our data. More independent modes need to be detected.

\subsubsection{Rotation of HS\,0507+0434B}

We use the frequency spacings within the three triplets to determine the
rotation period of HS\,0507+0434B. The average frequency separation
between the outer two components of the triplets is $7.48\pm0.07$ $\mu$Hz.
Assuming that all modes are $\ell=1$, that the star rotates uniformly, and
adopting first-order rotational splitting coefficients $C_{k,\ell}$ from
Brassard et al. (1992b, for masses around 0.6$M_{\sun}$ and radial
overtones corresponding to the observed period range), this leads to a
rotation period of $P_{\rm rot}=1.70\pm0.11$\,d. The main contribution to
the uncertainty in $P_{\rm rot}$ comes from the poorly-constrained
$C_{k,\ell}$. The dependence of $C_{k,\ell}$ on the radial overtone may be
one of the reasons why the triplets do not show exactly the same spacing.
Our estimate of the rotation period is perhaps the most reliable for any
cool ZZ Ceti star to date: consistent multiplet structure has never been
observed in these objects before.

We can also derive the inclination angle $\Theta$ from the relative
amplitudes within the triplets, but this requires several questionable
assumptions, the most important of which is equal physical amplitudes
within a multiplet. We mainly list the resulting $\Theta$ for
completeness: it would be $\Theta=78\degr\pm3\degr$, i.e. we would see the
star almost equator-on.

\subsubsection{Frequency differences and ratios}

Asymptotic theory (e.g. Tassoul 1980) predicts that the pulsation periods
of high-order g-modes, such as those excited in HS\,0507+0434B, should be
equally spaced. This is roughly true (allowing for other effects, e.g.
mode trapping to operate), but Table\,2 holds a number of surprises. The
modes $f_1, f_2$ and $f_5$, and hence $f_3$ and $f_6$ as well as $f_4$ and
$f_7$ are, within the errors of frequency determination, equally spaced in
{\it frequency}. The combination peak at the corresponding frequency
difference is also clearly detected.

The second interesting result from Table\,2 concerns the frequency ratios
of the ``fundamental'' modes: the ratio of $f_1$/$f_2$ is $0.750132 \pm
0.000018$ and the frequency ratio $f_2$/$f_5$ amounts to $0.800060 \pm
0.000011$, i.e. the frequencies of these modes relate almost exactly as
3:4:5! In addition, the frequency ratio of $f_7$/$f_8$ is $0.800254 \pm
0.000008$, very close to 4:5 as well, but there is no frequency difference
$f_7-f_8$ within the detected combination peaks. We found a combination
peak at $f_8-f_5$, but the frequency ratio of the corresponding parent
modes is $0.797543 \pm 0.000008$, which is not as close to the 3:4 and 4:5
ratios exhibited by the other modes.

Despite the close coincidence of these frequency ratios with
integer-number ratios, they are still different from the exact values by
several $\sigma$. However, we also keep in mind that the errors quoted in
Table 2 assume white noise and are therefore likely to be underestimates.
In particular, as we will argue in detail later, at least some of the
fundamental modes appear to exhibit temporal frequency or amplitude
variations, which increases our error estimates severely.

We checked whether it is justifiable to assume that exact integer-fraction
frequency ratios are present. Using {\tt Period98}, we fixed the
above-mentioned frequency ratios to exactly 3/4 or 4/5 and examined the
quality of such a fit with the Bayes Information Criterion (e.g. see
Handler et al. (2000) for a description) and by computing amplitude
spectra of the residuals. Both tests strongly suggested that exact 3/4 or
4/5 frequency ratios (between any of the mode groups) are not an
acceptable solution.

In any case, similar ``magic numbers'' were found among the frequencies of
other pulsating white dwarfs, e.g. the cool ZZ Ceti stars GD\,154
(Robinson et al. 1978), G\,191-16 (Vauclair et al. 1989) and BPM\,31594
(O'Donoghue, Warner \& Cropper 1992) or the pulsating DB white dwarf
PG\,1351+489 (Winget, Nather \& Hill 1987), but their agreement with
fractions of integers was much poorer than in the present case. The
majority of pulsating white dwarfs do not show integer-fraction frequency
ratios between their mode frequencies.

The explanation of these frequency ratios between the pulsation modes of
HS\,0507+0434B is not straightforward. We speculate that resonances could
be involved; this could reproduce the observed frequency ratios (e.g. see
Buchler, Goupil \& Hansen 1997). However, to our knowledge, the
possibility of 3:4 or 4:5 resonances in pulsating stars has not yet been
explored, perhaps because the coupling is expected to be weak (Moskalik,
private communication).

\subsection{The meaning of the remaining signals}

Having finished the discussion of the ten fundamental frequencies, we will
now try to explain the signals in the light curves of HS\,0507+0434B
listed at the bottom of Table 2.

The frequency $f_{49}$ may correspond to a further independent mode. After
prewhitening of the ten fundamental frequencies and their combinations, it
is the highest peak in the residuals, and it is not explicable with a
reasonable combination frequency. Furthermore, some of the highest peaks
in the residual amplitude spectrum ($f_{51}-f_{54}$) can be explained by
combinations of $f_{49}$ and one of the ten fundamental frequencies.

After prewhitening of $f_{49}$, a peak separated by $14.22 \pm 0.17 \mu$Hz
($f_{50}$) remains in the residuals. Assuming that our $\ell=1$
identifications for $f_2-f_{10}$ are correct, $f_{49}$ cannot be $\ell=1$,
as it is too close to the $f_2-f_4$ triplet. If $f_{49}$ was part of an
$\ell=2$ quintuplet, then $f_{50}$ can be explained as another
rotationally split component of it: the separation between modes of
$|\Delta m|=2$ would be 12.5 $\mu$Hz, which is consistent with the
observed spacing as it could be blended with the 1~c/d alias of $f_{49}$.

The previous hypothesis does not, however, account for the number of peaks
in the 3500$\mu$Hz region in the lowest panel of Fig.\,5, assuming they
are all combinations of $f_{49}$ or $f_{50}$ with $f_{2}-f_{4}$. It is
interesting to note that $\ell=1$ pulsation modes around frequencies of
3500$\mu$Hz are frequently present in pulsating DA white dwarfs (Kleinman
1999). The power in this frequency region could therefore be due to such a
mode, due to combination frequencies or due to a combination of both. We
can therefore not determine whether the signals around 1700$\mu$Hz are
combinations of such modes with $f_{2}-f_{4}$ or represent independent
modes.

We now turn to the signals at $f_{55}$ and $f_{56}$, which are located in
the same frequency domain as the modes $f_{2}-f_{4}$. It is immediately
obvious that they cannot be independent pulsation modes of $\ell=1$ or
$\ell=2$, as their frequencies do not result in consistent multiplet
structure. We suggest that these peaks are manifestations of amplitude
and/or frequency variations of $f_{2}-f_{4}$. $f_{56}$ is unresolved from
$f_{2}$ and $f_{55}$ is unresolved from the 1-day alias of $f_{4}$;
$f_{2}$ and $f_{4}$ are the strongest components of this multiplet.
Regrettably, our data set is too small to investigate the temporal
behaviour of the $f_{2}-f_{4}$ multiplet in detail. Some support for this
interpretation comes from the combination frequencies of $f_{2}-f_{4}$, as
signals suggestive of these temporal variations are also present in the
corresponding frequency range.

Temporal variability of the pulsation spectra of cool ZZ Ceti stars is not
uncommon; it is rather the rule than the exception (see Kleinman 1995),
and it is consistent with theory (e.g. Wu \& Goldreich 2001). Therefore
the presence of $f_{55}$ and $f_{56}$ does not lead us to doubt our mode
identification. On a longer time scale, however, the amplitudes of
HS\,0507+0434B seem remarkably stable compared to other cool ZZ Ceti
stars: the results from the discovery observations (Jordan et al. 1998)
and spectrophotometric light curves (Kotak \& van Kerkwijk 2001), both
obtained years apart from each other and from our measurements, seem to be
adequately reproduced by our frequency solution. There is also no evidence
for temporal variations of the $f_5-f_7$ and $f_8-f_{10}$ triplets.

\subsection{Combination frequencies}

Combination frequencies potentially contain a wealth of astrophysical
information, which we are only just beginning to explore (e.g. see Wu 1998
and Vuille 1999). As a further step in this direction, we will make a
detailed comparison of our observations with the model by Wu (1998, 2001,
hereinafter ``Wu's model'')\footnote{We have also tested the model by
Brassard, Fontaine \& Wesemael (1995), but found it to underestimate the
observed combination frequency amplitudes by more than an order of
magnitude. We will therefore not discuss it further.}. This model is based
on Brickhill's (1992) argument that the combination frequencies are
generated by the surface convection zone as the eigenmodes pass through
them (see Vuille 2000 for a summary). The combination frequencies are thus
a manifestation of the distortion of the light curve generated by the
convection zone, which means that a quantification of these distortions
can be used to infer physical properties of the convection zone. These
can, in turn, yield other constraints, like the relative temperatures of
ZZ Ceti stars over their instability strip. The form of some of the
distortions also depends on the spherical degree $\ell$ and azimuthal
order $m$ of their parent modes, which holds the promise of achieving a
mode identification for the parents by making use of their combinations.

HS\,0507+0434B is ideally suited for the application of Wu's model. Due to
the triplet structures observed within the pulsation modes, assignments of
$\ell, m$ are possible. This is a definite asset for combination frequency
study, which we will take advantage of in what follows.

Wu's model provides analytical formulae which bring the observable
parameters in direct connection with the theoretical quantities to be
inferred. It can therefore be immediately applied without the use of
specific model calculations. However, the model is built on a perturbation
analysis, which means it can only be used within certain limits of the
pulsational amplitude of the star. Ising \& Koester (2001) examined these
limits by means of numerical model calculations and found that it is only
valid for variations with photometric amplitudes smaller than about 4\% (a
necessary but not sufficient condition). All the modes of HS\,0507+0434B
detected by us have amplitudes smaller than that.

In what follows, we will apply Wu's model step-by-step to assess its
potential. We rewrite Wu's equations so that the results from the
frequency analysis of the photometric data can be directly used as input.
We restrict ourselves to the second-order combinations, which are the most
numerous and have the best $S/N$ of all the combination frequencies.

\subsubsection{Convective thermal time $\tau_c$}

This is the first parameter of Wu's model to be tested. It can be inferred
from the amplitudes and frequencies of combinations where both the
difference and sum frequencies are observed. We rewrite Eq.\,21 from Wu
(2001) as:
\begin{equation}
\tau_c=\frac{1}{2\pi \left|f^2_i-f^2_j\right|}\sqrt{\frac{[r
A_{i-j}(f_i+f_j)]^2-[A_{i+j}(f_i-f_j)]^2}{A_{i+j}^2-r^2 A_{i-j}^2}},
\end{equation}
where
\[
r \equiv \frac{c(m_i,m_j)}{c(m_i,-m_j)},
\]
where $A_{i-j}$ is the amplitude of the difference frequency, $A_{i+j}$ is
the amplitude of the sum frequency, $f_i$ and $f_j$ are the frequencies of
the parent modes, and values for $c(m_i,m_j)$, a constant depending on the
values of the azimuthal order $m$ of the parent modes, are listed in
Table\,3. The amplitudes can be given in any (consistent) units, but the
frequencies need to be in Hz for $\tau_c$ to result in units of seconds.

\begin{table}
\caption[]{$m$-dependent constants for determining the amplitudes of a
combination peak, adapted from Wu (2001). $\Theta$ is the inclination of the
star's pulsational axis to the line of sight. The results are for
$\ell=1$.}
\begin{center}
\begin{tabular}{ccc}
\hline
$m_1$ & $m_2$ & $c_{(m_i,m_j)}$\\
\hline
0 & 0 & 0.65 +0.45/cos$^2 \Theta$\\
0 & +1 & 0.65\\
0 & $-$1 & 0.65 \\
+1 & +1 & 0.65 \\
+1 & $-$1 & 0.65 +0.90/sin$^2 \Theta$ \\
$-$1 & $-$1 & 0.65 \\
\hline
\end{tabular}
\end{center}
\end{table}

The application of Eq.\,1 requires a mode identification. This means we
have to assume the three triplets are rotationally split m-components of
$\ell=1$ modes. Consequently, we used the five mode pairs $f_4$/$f_5$,
$f_2$/$f_8$, $f_4$/$f_8$, $f_5$/$f_8$ and $f_2$/$f_5$ for which Eq.\,1 can
be applied. Whereas the combinations involving $f_8$ implied a range of
$\Theta>45\degr$ and 60\,s $<\tau_c<$ 120\,s, the other two mode pairs
gave unphysical results. Thus we refrain from determining $\tau_c$ in this
way.

An alternative possibility to determine $\tau_c$ uses the observed phase
differences of the combination frequencies relative to those of the parent
modes. This diagnostic is independent of any assumed mode identification.
We have therefore calculated these parameters from our observations for
the second-order combinations; these are listed in Table 4 together with
other parameters required later on.

\begin{table}
\caption[]{Amplitude ratios and phase differences of the two-mode
combinations relative to their parent modes. Signals indicated with a
asterisk ($\ast$) were used for determining $\tau_c$, those marked with a
diamond ($\diamondsuit$) were utilized to determine $|2\beta+\gamma|$;
their amplitudes are independent of the inclination angle $\Theta$.
Frequencies denoted with a plus sign ($+$) were also used to derive
$|2\beta+\gamma|$, but their relative amplitudes depend on $\Theta$. $A_{i
\pm j}$ is the amplitude of the combination peak of the modes with
frequencies $f_i$ and $f_j$, and $A_i$ and $A_j$ are the corresponding
amplitudes of the parents. Similarly, $\phi_{i \pm j}$ is the phase of the
combination of modes $f_i$ and $f_j$, and $\phi_i$ and $\phi_j$ are phases
of the parents. $f_{51}$ to $f_{54}$ are only listed for completeness.
Note however that their amplitude ratios are larger than those of the
combinations of the other modes.}
\begin{center}
\begin{tabular}{lcc}
\hline
ID & $A_{i \pm j}/A_iA_j$ & $\Delta\phi=\phi_{i \pm j}-(\phi_i \pm \phi_j)$\\
 & (mag)$^{-1}$ & (rad) \\
\hline
$f_{11}$$^{\ast}$ & 6.9 $\pm$ 1.8 & 0.35 $\pm$ 0.26 \\
$f_{12}$$^{\ast +}$ & 23.9 $\pm$ 1.0 & -0.03 $\pm$ 0.04 \\
$f_{13}$$^{\ast\diamondsuit}$ & 9.6 $\pm$ 1.9 & 0.13 $\pm$ 0.20 \\
$f_{14}$$^{\ast\diamondsuit}$ & 5.9 $\pm$ 0.8 & 0.52 $\pm$ 0.14 \\
$f_{15}$$^{\ast\diamondsuit}$ & 4.8 $\pm$ 1.0 & 0.77 $\pm$ 0.21 \\
$f_{16}$ & 17.2 $\pm$ 2.3 & -0.35 $\pm$ 0.14 \\
$f_{17}$ & 14.3 $\pm$ 1.0 & 0.39 $\pm$ 0.07 \\
$f_{18}$$^{\ast}$ & 8.9 $\pm$ 1.3 & 0.12 $\pm$ 0.14 \\
$f_{19}$$^{\ast +}$ & 14.8 $\pm$ 1.4 & 0.89 $\pm$ 0.10 \\
$f_{20}$$^{\ast\diamondsuit}$ & 6.6 $\pm$ 0.6 & 0.52 $\pm$ 0.09 \\
$f_{21}$ & 13.1 $\pm$ 0.6 & 1.40 $\pm$ 0.05 \\
$f_{22}$$^{\ast\diamondsuit}$ & 6.9 $\pm$ 0.7 & 0.62 $\pm$ 0.10 \\
$f_{23}$ & 20.8 $\pm$ 0.9 & 0.70 $\pm$ 0.04 \\
$f_{24}$$^{\ast\diamondsuit}$ & 3.1 $\pm$ 0.4 & 0.24 $\pm$ 0.13 \\
$f_{25}$$^{\ast +}$ & 11.8 $\pm$ 1.0 & 0.54 $\pm$ 0.08 \\
$f_{26}$$^{\ast\diamondsuit}$ & 9.6 $\pm$ 2.2 & 0.10 $\pm$ 0.23 \\
$f_{27}$$^{\ast +}$ & 9.4 $\pm$ 1.0 & 0.16 $\pm$ 0.11\\
$f_{28}$ & 10.9 $\pm$ 1.0 & 0.93 $\pm$ 0.09 \\
$f_{29}$ & 5.0 $\pm$ 0.7 & 1.01 $\pm$ 0.14 \\
$f_{30}$$^{\ast +}$ & 3.3 $\pm$ 0.6 & 0.78 $\pm$ 0.17 \\
$f_{31}$ & 5.6 $\pm$ 0.6 & 2.33 $\pm$ 0.10 \\
$f_{32}$$^{\ast}$ & 17.4 $\pm$ 1.3 & 1.23 $\pm$ 0.08 \\
$f_{51}$ & 27.1 $\pm$ 4.3 & 2.67 $\pm$ 0.16\\
$f_{52}$ & 38.5 $\pm$ 9.2 & -2.97 $\pm$ 0.24\\
$f_{53}$ & 32.0 $\pm$ 5.7 & 0.28 $\pm$ 0.18 \\
$f_{54}$ & 26.2 $\pm$ 6.4 & -0.01 $\pm$ 0.25 \\
\hline
\end{tabular}
\end{center}
\end{table}

We can now estimate $\tau_c$ using Eq.\,15 of Wu (2001), for which we find
the following form most practical:
\begin{equation}
\Delta\phi= \pi/2-\arctan(2\pi(f_i \pm f_j)\tau_c),
\end{equation}
where the meaning of $\Delta\phi$ is given in the caption of Table 4, and
the other parameters were explained before (Eq.\,1).

As some of the combination signals have more than one potential set of
parent modes due to degeneracy of their frequencies, we only used
combinations which have unique parents or where one pair of parents
strongly dominates possible others. Thus we only included combinations
where the product of the amplitudes of one pair of parents exceeds that of
all potential others by at least a factor of 4. These signals are
indicated with an asterisk in Table 4. The application of Eq.\,2 does not
require a mode identification.

In Fig.\,6, the measured values of $\Delta\phi$ are compared to the
predictions of Eq.\,2. Theory and observations are roughly consistent. A
best fit to the observed phase differences results in $\tau_c \approx
110$\,s, but every $\tau_c$ between 80 and 150\,s seems to give an
acceptable fit.

In the framework of the same convection model, the condition $2\pi
f_i\tau_c>1$ needs to be fulfilled for pulsation modes to be driven
(Goldreich \& Wu 1999). The best-fit $\tau_c=110$\,s yields $2\pi
f_1\tau_c=0.93$, marginally consistent with these stability
considerations. It also becomes clear that reliable results for the
difference frequencies are vital for determinations of $\tau_c$.

\begin{figure}
\includegraphics[width=99mm,viewport=-60 -20 367 218]{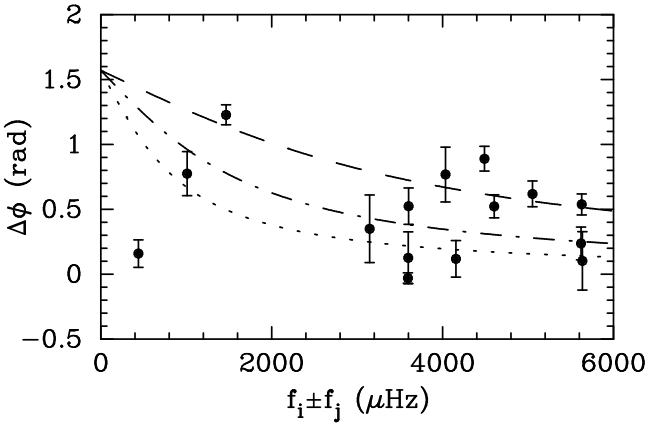}
\caption[]{A comparison of the relative phases of the two-mode combination
frequencies (filled circles with error bars) with theoretical predictions for
different values of $\tau_c$ by means of Eq.\,2. The dashed line is for
$\tau_c=50$\,s, the dashed-dotted one for $\tau_c=110$\,s (best fit) and
the dotted line for $\tau_c=200$\,s. Except for one outlier, the
theoretically predicted trend is roughly followed, but an accurate
determination of $\tau_c$ is not possible.}
\end{figure}

Unfortunately, the low frequencies of difference signals makes their
amplitudes and phases observationally harder to detect, as they are more
affected by variations in sky transparency and extinction. The use of CCD
detectors, such as in our case, helps in this respect, but does not fully
eliminate this problem. From the theoretical point of view, Wu (2001)
noted that the combinations of lowest frequency are also the least
reliable, which may be caused by the assumption of a dynamically and
thermally coherent convection zone breaking down. This may explain the
discrepancies we have noted, but we conservatively do not reject certain
measurements based on these arguments.

\subsubsection{$|2\beta+\gamma|$ and inclination angle $\Theta$}

The parameter $|2\beta+\gamma|$ describes the response of the stellar
material to the pulsations, and $\Theta$ is the inclination of the star's
pulsational axis to the line of sight, which is assumed to coincide with
the stellar rotation axis.

To determine these parameters, we rewrite Eq.\,20 of Wu (2001) as:
\begin{equation} c_{(m_i,m_j)}|2\beta+\gamma|=1.086\frac{A_{i \pm
j}}{A_iA_j} \frac{4\alpha_{\rm V}\sqrt{1+(2\pi(f_i \pm
f_j)\tau_c)^2}}{n_{ij}{2\pi(f_i \pm f_j)\tau_c}} \end{equation} where
$A_{i \pm j}$ is the amplitude of the combination peak at frequency $f_i
\pm f_j$, $A_i$ and $A_j$ are the amplitudes of the parent modes, 1.086 is
the scale factor between magnitude and intensity variations,
$c_{(m_i,m_j)}$ was again taken from Table 3, $n_{ij}$ is the number of
possible permutations of $i$ and $j$ ($n_{ij}=2$ if $i\neq j$, $n_{ij}=1$
otherwise), theoretical values of $|2\beta+\gamma|$ are around 15, see Wu
(2001), $\tau_c$ is again the convective thermal time and $\alpha_{\rm V}$
is a scale factor from bolometric luminosity variations to those in the
observational passband (our data can be considered as being obtained in a
``wide V filter'', which leads us to adopt $\alpha_{\rm V}=0.72$). The
amplitudes used in Eq.\,3 need to be in magnitudes, the frequencies are in
units of Hz, and $\tau_c$ is in seconds.  

We can now constrain $|2\beta+\gamma|$. We start by applying
Eq.\,3 to the combinations marked with diamonds ($\Theta$-independent
$c_{(m_i,m_j)}$-values) and plus signs ($\Theta$-dependent
$c_{(m_i,m_j)}$-values) in Table 4. These combinations have, as
previously, a unique or at least one dominant pair of parents, and we
assign an $m$ value to these parents by assuming all triplets are
generated by rotationally split $\ell=1$ modes. 

The mean value of $|2\beta+\gamma|$ from the $\Theta$-independent
combinations is $4.6\pm 0.7$, lower than predicted by theory. The
$\Theta$-dependent combinations show a mean $c_{(m_i,m_j)}|2\beta+\gamma|$
of $8.3\pm2.1$. They are all combinations of parents identified with
$(m_i,m_j)=+1,-1$. Assuming $|2\beta+\gamma|=4.6$, we then obtain
$\Theta>48\degr$. We show the individual values of $|2\beta+\gamma|$ for 
$\Theta=80\degr$ in Fig.\,7.

\begin{figure}
\includegraphics[width=99mm,viewport=3 8 328 190]{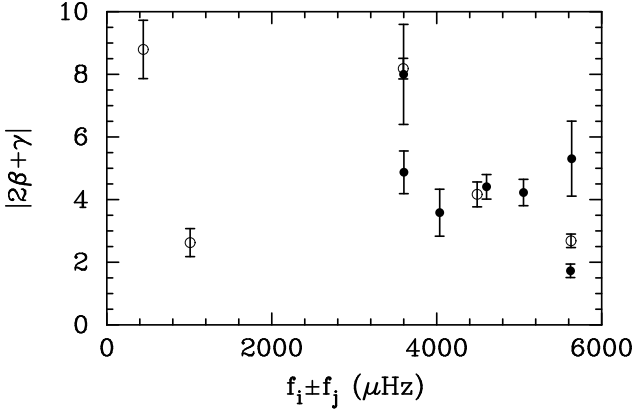}
\caption[]{Values of $|2\beta+\gamma|$ from the relative amplitudes of the
two-mode combinations (Eq.\,3), assuming $\Theta=80\degr$. Filled circles
are for combinations whose $c_{(m_i,m_j)}$ is independent of the
inclination angle $\Theta$; open circles denote $\Theta$-dependent 
combinations.}
\end{figure}

\subsubsection{Theoretically predicted amplitude spectra}

We now check how well Wu's model can reproduce the amplitude spectrum of
the second-order combination frequencies. We used our best frequency
solution (Table 2), and took the frequencies, amplitudes and phases of all
detected second-order combinations to create a synthetic light curve
(hereinafter SLC), sampled in exactly the same way as the observational
data, to which the model results were compared. This was done for clarity
of presentation, as it disregards the observational noise. However, the
basic results including noise are the same.

We then used Wu's model to predict the amplitudes and phases of the
combination frequencies, varying the three free model parameters within
reasonable limits (80\,s $< \tau_c <$ 150\,s, 5 $<|2\beta+\gamma|<$ 25,
and $\epsilon <\Theta< \pi/2-\epsilon$ for $\epsilon>0$) and computed the
amplitude spectra of these theoretical light curves. We did not compare
the theoretically predicted parameters directly to those of the SLC as our
data set has finite temporal resolution and gaps, which will cause some of
the computed signals to interact either through their aliases, or because
some will not be resolved.

We attempted to find amplitude spectra which resembled that of the SLC
closely. We computed a series of theoretical light curves varying the
parameters in the intervals given above, calculated their amplitude
spectra, subtracted those from the amplitude spectrum of the SLC and used
the resulting residual scatter as a criterion for the adoption of a set of
parameters. Values of $|2\beta+\gamma|=10.0\pm1.5$, $80 < \tau_c <$ 110\,s
and $\Theta=80\pm2\degr$ resulted in the best representation of the SLC.
Disregarding the combination frequency differences because of the reasons
given at the end of Sect. 4.3.1 does not change the results significantly;
best-fit parameters would be $|2\beta+\gamma|=11.3\pm1.0$, $80 < \tau_c <$
110\,s and $\Theta=79\pm1\degr$. We show the observed amplitude spectrum
of the SLC and a theoretical one which resembles it well in Fig.\,8.

\begin{figure*}
\includegraphics[width=99mm,viewport=55 -80 375 125]{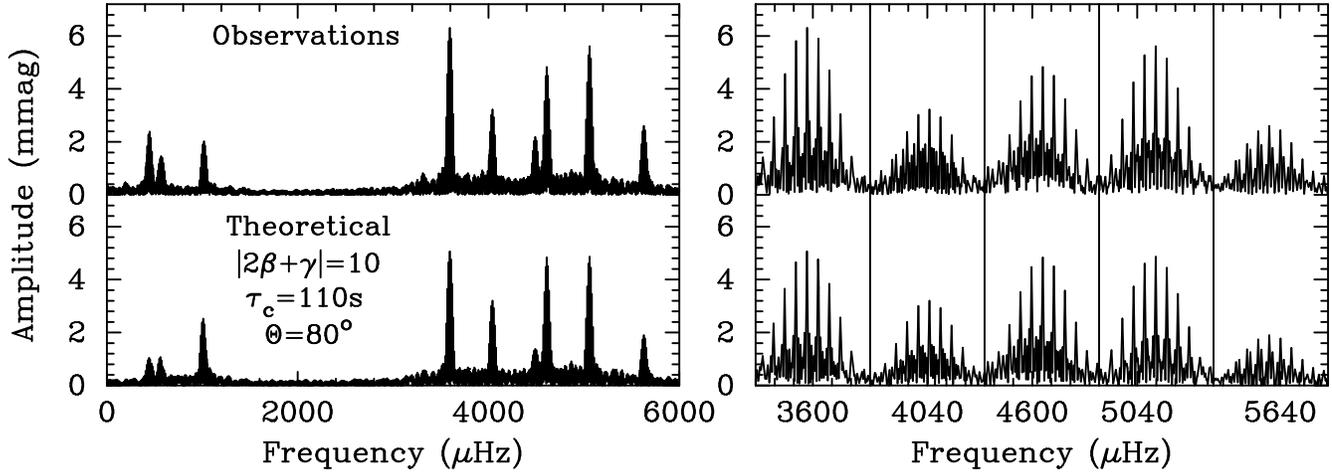}
\caption[]{Upper half: the amplitude spectrum of the two-mode combinations
in the SLC of HS\,0507+0434B. Lower half:  theoretical amplitude spectrum
computed with Wu's model and the parameters given in the left half. The
panels on the left side show the predicted amplitude spectrum globally,
whereas those on the right side present zoom-ins on the five strongest sum
frequencies.}
\end{figure*}

The left-hand panels of Fig.\,8 show the overall amplitude spectra of the
SLC and the theoretical second-order combination light curve. The global
agreement is quite good. In the right-hand panels of Fig.\,8 we show the
dominant structures with better resolution. The inter-multiplet structures
also seem well reproduced.

\subsubsection{Direct light-curve fitting}

The previous comparison did not make full use of the phase information
contained in the light curves and predicted by Wu's model. As a final test
we therefore compared the SLC and the theoretically predicted combination
light curve directly. We again computed theoretical light curves for a
wide set of free parameters and evaluated the rms residual between the SLC
and theoretical light curves as our criterion.

The result of this test was very similar to that of the previous one. The
best fits were obtained with the following parameter values:
$|2\beta+\gamma|=9.7\pm1.0$, $80 < \tau_c <$ 110\,s and
$\Theta=79\pm1.5\degr$, and $|2\beta+\gamma|=11.7\pm1.0$, $80 < \tau_c <$
110\,s, $\Theta=79\pm1.0\degr$ disregarding the frequency differences.  
These light-curve fits resembled the observations well, but not as well as
our frequency solution.

We conclude from our applications of Wu's model that it is possible to
obtain plausible values for the convective thermal time $\tau_c$, but
with large uncertainties. Our results for $|2\beta+\gamma|$ also appear
reasonable, but the small values obtained in Sect. 4.3.2 suggest caution.
The parameter which seems to be best constrained is the inclination angle
$\Theta$; its preferred values were around 80\degr. This is consistent
with the $\Theta$-determination from the relative amplitudes of the
components of the three mode triplets (Sect. 4.1.3).

We do not want to overemphasize this consistency, as we believe the
results on the value of $\Theta$ from Sect. 4.1.3 and that of the
application of Wu's model are not fully independent. Hence, we do not dare
to base further interpretations on this result.

The results from Sect. 4.3.1 and 4.3.2 are somewhat different from those
of Sect. 4.3.3 and 4.3.4. This may be due to the stronger influence of the
combination frequency differences on the general results in the first two 
investigations. We think that the outcome of Sect. 4.3.3 and 4.3.4 is more 
meaningful as it is dominated by the best-determined frequency sums and 
because we have used more of the information in the observed light curve.

\subsubsection{Application to $\ell=2$ modes}

Having shown that Wu's model yields consistent results under the
assumption that all the triplet components detected in Sect. 3 are due to
modes of $\ell=1$, we now examine whether we can reproduce the
observations assuming the triplets are caused by $\ell=2$ modes. To this
end, we first computed the $c_{(m_i,m_j)}$ for the $\ell=2$ case, and we
list them in Table 5.

\begin{table}
\caption[]{Same as Table~3, but for $\ell=2$ modes}
\begin{center}
%\scriptsize
\begin{tabular}{>{$}r<{$}>{$}r<{$}>{$}c<{$}}
\hline
m_1 & m_2 & c_{(m_i,m_j)}\\
\hline
2 & 2 & -0.197 \\ 
1 & 2 & -0.197 \\ 
0 & 2 & \frac{
       -3.939 - 0.592\,\cos (2\,{{\Theta }})}{1 + 
     3\,\cos (2\,{{\Theta }})} \\ 
-1 & 2 & -0.197 + 
   1.871\,/ \sin^2 {{\Theta }} \\ 
-2 & 2 & \left( 4.457 - 
     1.772\,\cos (2\,{{\Theta }}) - 
     0.02465\,\cos (4\,{{\Theta }}) \right)
   /{\sin^4 {{\Theta }}} \\ 
% 2 & 1 & -0.197 \\ 
1 & 1 & 
   \left( 0.369 - 0.0986\,\cos (2\,{{\Theta }}) \right)
   /{\cos^2 {{\Theta }}} \\ 
0 & 1 & \frac{1.674 - 
     0.592\,\cos (2\,{{\Theta }})}{1 + 3\,\cos (2\,{{\Theta }})}
   \\ 
-1 & 1 & \frac{ 1.342 + 0.234\,\cos (2\,{{\Theta }}) + 
     0.02465\,\cos (4\,{{\Theta }})}
    {({\sin ({{\Theta }})}\,{\cos ({{\Theta }})}^2} \\ 
-2 & 1 & 
     -0.197 + 1.871/{\sin^2 {{\Theta }}} \\ 
0 & 0 & \frac{33.584 + 10.042\,\cos (2\,{{\Theta }}) - 
     0.8876\,\cos (4\,{{\Theta }})}{{\left( 1 + 
        3\,\cos (2\,{{\Theta }}) \right) }^2} \\ 
-1 & 0 & \frac{
      1.674 - 0.592\,\cos (2\,{{\Theta }})}{1 + 
     3\,\cos (2\,{{\Theta }})} \\ 
-2 & 0 & \frac{-3.939 - 
     0.592\,\cos (2\,{{\Theta }})}{1 + 3\,\cos (2\,{{\Theta }})}
   \\ 
-1 & -1 & \left( 0.369 - 
     0.0986\,\cos (2\,{{\Theta }}) \right) 
   /{\cos^2 {{\Theta }}} \\ 
-2 & -1 & -0.197 \\ 
-2 & -2 & -0.197 \\
\hline
\end{tabular}
\end{center}
\end{table}

We applied direct light-curve fitting (Sect.  4.3.4) as it uses most of
the light curve information and as its consistency has already been
demonstrated. We performed the fits for all 28 possible $m$ assignments to
the triplet components for $\ell=2$, and we did so with and without
considering the combination frequency differences.

Our best fits were obtained for $m$ values of $(-1,0,1)$ for all three
triplets and the parameters $|2\beta+\gamma|=3.1$, $\tau_c =$ 90\,s and
$\Theta=46\degr$, which produced residuals 3\% higher than that for the
best fit with $\ell=1$ combinations. Disregarding the difference 
frequencies, the best fits resulted from the same $m$ assignments as above
and $|2\beta+\gamma|=3.5$, $\tau_c =$ 110\,s and $\Theta=46\degr$. The
corresponding residuals were 20\% higher than those for the best $\ell=1$
solution. Only those fits with $m$ assignments of either $(-1,0,1)$ or
$(-2,0,2)$ gave nearly as good results as those for the $\ell=1$ case.

However, the best-fit values for $|2\beta+\gamma|$ are much smaller than
predicted by theory and thus appear unreasonable. This is easy to
understand (e.g. see Wu 2001) as $\ell=2$ modes suffer a larger amount of
geometrical cancellation (Dziembowski 1977), and thus the photometric
amplitudes of their combinations are higher than those of the combinations
of $\ell=1$ modes.

The combination of the small $|2\beta+\gamma|$ values with the poorer fits
that resulted despite the larger (by a factor of 28) number of
possibilities for matches provided by possible $\ell=2$ modes, leads us to
suggest that the three triplets detected in Sect. 3 are indeed pulsation
modes of $\ell=1$.

As noted in Sect. 4.1.1, it may be possible that one or the other $\ell=2$
mode disguised itself in a predominant $\ell=1$ pattern, which cannot be
confirmed or disproven by the analysis of the combination frequencies, but 
again we believe such an interpretation would be contrived.

\subsubsection{Mode identification for $f_1$ by means of its combination 
frequencies?}

Having obtained support for the $\ell=1$ for the three triplets, it is now 
tempting to try a mode identification for the singlet $f_1$. 
Unfortunately, only three two-mode combinations of $f_1$ are observed, 
which is less than expected considering its amplitude. Indeed, the 
light-curve fitting did not result in acceptable solutions for any $m$ 
value assuming $f_1$ was due to an $\ell=1$ mode; more combination signals 
than actually observed should be present.

\begin{table}
\caption[]{Same as Table~3, but for modes where $\ell_1=1$ and
$\ell_2=2$.}
\begin{center}
%\scriptsize
\begin{tabular}{>{$}r<{$}>{$}r<{$}>{$}c<{$}}
\hline
m_1 & m_2 & c_{(m_i,m_j)}\\
\hline
1 & 2 & 0.271\\
 0 & 2 & 0.271\\
 -1 & 2 & 0.271+ 2.244/\sin^2\theta \\
 1 & 1 & 0.271\\
 0 & 1 & 0.271+ 0.561/\cos^2\theta \\
 -1 & 1 & 0.271+ 1.122/\sin^2\theta \\
 1 & 0 & \frac{-1.973 + 0.814\,\cos (2\,\theta )}{1 + 3\,\cos (2\,\theta )} \\
 0 & 0 & \frac{ 4.760 + 0.814\,\cos (2\,\theta )}{1 + 3\,\cos (2\,\theta )} \\
-1 & 0 & \frac{-1.973 + 0.814\,\cos (2\,\theta )} {1 + 3\,\cos (2\,\theta )} \\
 1 & -1 & 0.271+ 1.122/\sin^2\theta \\
 0 & -1 & 0.271+ 0.561/\cos^2\theta \\
 -1 & -1 & 0.271\\
 1 & -2 & 0.271+ 2.244/\sin^2\theta \\
 0 & -2 & 0.271\\
 -1 & -2 & 0.271\\
\hline
\end{tabular}
\end{center}
\end{table}

We are left with the possibility that $f_1$ is due to a mode of $\ell=2$.  
Consequently, we computed $c_{(m_i,m_j)}$ constants for combinations of
$\ell=1$ and $\ell=2$ modes, which we list in Table 6. The examination of
this hypothesis with the direct light curve fitting method yielded
inconclusive results, although the solutions were generally better as
under the $\ell=1$ assumption. This is not surprising as the
$c_{(m_i,m_j)}$ coefficients in Table 6 are generally smaller than those
in Table 3. A mode identification of $f_1$ by means of its combination
frequencies is however not possible at this point.

\section{Summary and Conclusions}

We obtained one week of time-resolved CCD photometry of the cool ZZ Ceti
star HS\,0507+0434B. We were able to detect ten ``fundamental
frequencies'', 38 combinations of these, and eight more signals. This
frequency solution explains the observed light variations satisfactorily.
Even though light curves of multi-periodic variables, such as pulsating
white dwarf stars, usually can only be deciphered with multi-site
observations, we believe in this case we were fortunate enough to achieve
just that.

The fundamental frequencies we detected consist of one singlet and three
almost-equally spaced triplets. The frequency spacings within the
individual triplets are very similar, which leads us to suggest they are
caused by rotational $m$-mode splitting. This is the first clear detection
of rotationally split triplets in a cool ZZ Ceti star. Assuming they are
due to $\ell=1$ modes, we infer a rotation period of 1.70$\pm$0.11\,d for
HS\,0507+0434B.

The periods of the three triplets support the identification with $\ell=1$
modes by comparison with other cool ZZ Ceti stars. The star's pulsational
spectrum determined by us is too poor for a detailed asteroseismological
analysis.

We reported the curious detection that three of the basic four structures
(one singlet, three triplets) are exactly equally spaced in frequency, not
in period as predicted by asymptotic theory. In addition, these three
structures show frequency ratios very close to (within 10$^{-4}$), but
still significantly different from, 3:4:5. The fourth basic structure also shows
a frequency ratio close to 4:5 with its neighbour. We speculate that
resonances, perhaps together with mode trapping, are responsible for these
frequency ratios.

Among the features in the star's pulsation spectrum which could not be
easily explained are two more signals, together with some of the
combinations of one with other mode frequencies. These frequencies could
correspond to components of an $\ell=2$ multiplet. We have also found
evidence for temporal amplitude and/or frequency variations of the
longest-period ``fundamental'' triplet.

We used the detected combination frequencies to test the predictions of
Wu's (1998, 2001) model. We were able to arrive at a believable-looking
estimate of the convective thermal time, but the result has large
uncertainties. Our determination of the inclination angle $\Theta$ appears
plausible. Theoretical predictions of the observed amplitude spectra of
the combination signals were reasonably successful as were attempts to fit
the observed light curves directly. They also supported the identification
of the three triplets with $\ell=1$.

We believe that the application of Wu's model is promising, but requires
further testing. This could for instance be done by using well-resolved
WET data or by confrontation with numerical models such as those by Ising
\& Koester (2001), in particular if the latter could be extended to
multi-periodic pulsations. A comparison of the numerical computations with
observations would then become very interesting.

\section*{ACKNOWLEDGMENTS}

We are indebted to the referee, Yanqin Wu, for her insightful report on
the manuscript which lead to considerable improvement of this paper. GH
thanks Noel Dolez for drawing his attention to this star, Steve Kawaler
and Chris Koen for useful discussions and Scot Kleinman for supplying his
{\tt lincom} code which was essential for this analysis. The constructive
and critical comments of Atsuko Nitta on an early draft version of this
paper are appreciated. ERC would like to thank S. Potter, D. Romero and E.
Colmenero for their support. MHM thanks D. E. Winget and A. Mukadam for
useful discussions.

\bsp

\end{document}